\documentclass[%
preprint,
 amsmath,amssymb,
 aps,
]{revtex4-2}

\usepackage{adjustbox}
\usepackage{physics}
\usepackage{amsmath}
\usepackage{array}
\usepackage{setspace}
\newcolumntype{C}[1]{>{\centering\arraybackslash}m{#1}}
\usepackage{xcolor}
\usepackage{graphicx}
\usepackage{dcolumn}
\usepackage{bm}
\usepackage{caption}
\usepackage{subcaption}
\usepackage{rotating}
\usepackage{subcaption}

\begin{document}

\preprint{APS/123-QED}

\title{Efficient Deterministic-Stochastic Representation of the Coulomb Operator in Real Space}

\author{Nafiz Faiaz}
\affiliation{Department of Chemistry and Biochemistry, University of California, Los Angeles,
Los Angeles, CA, 90095, USA}
\author{Tucker Allen}
\affiliation{Department of Chemistry and Biochemistry, University of California, Los Angeles,
Los Angeles, CA, 90095, USA}
\author{Dimitri Bazile}
\affiliation{Department of Chemistry and Biochemistry, University of California, Los Angeles,
Los Angeles, CA, 90095, USA}
\author{Kajsa Williams}
\affiliation{Department of Chemistry and Biochemistry, University of California, Los Angeles,
Los Angeles, CA, 90095, USA}
\author{Daniel Neuhauser}
\affiliation{Department of Chemistry and Biochemistry, University of California, Los Angeles,
Los Angeles, CA, 90095, USA}

\begin{abstract}
We present an efficient mixed deterministic-stochastic approach for constructing the Coulomb operator in real space that preserves accuracy across the full spectral range. The dominant long-range components of the interaction are captured deterministically via a compact, low-rank approximation, constructed using Chebyshev-filtered subspace iteration, while the remaining spectral tail is treated using unbiased probing with a small number of stochastic vectors. The resulting operator is benchmarked within self-consistent field calculations for the extended Hubbard Hamiltonian on periodic, perturbed, and non-periodic three-dimensional lattices. We account for stochastic bias in observables by implementing a jackknife correction. The method systematically improves with deterministic rank and stochastic sample count, while substantially reducing the computational cost of constructing the Coulomb matrix.

\end{abstract}

\maketitle

\section{Introduction}

The Coulomb potential operator is a central component of electronic structure theory. The electron-electron interaction is long-ranged, producing dense and ill-conditioned matrices whose direct application scales quadratically with system size. This challenge is particularly severe in real-space treatments. While reciprocal-space formulations can exploit convolution structure and FFTs to reduce the application cost to $\mathcal{O}(N\log N)$  \cite{martin2004electronic}, a real-space grid discretization, in which position space is represented by $N$ grid points, exposes the full $\mathcal{O}(N^{2})$ scaling. This bottleneck is prominent in several electronic-structure methods, such as density functional theory (DFT) and Hartree-Fock (HF) theory, where the application of Coulomb and Fock exchange terms dominates the computational cost of calculations \cite{LinearExact, Ewald_particle_mesh, planewave}. Efficient representations of the Coulomb operator are therefore useful for reducing this cost in large real-space calculations. 

Substantial efforts have been devoted to reducing the cost of Coulomb interactions in real-space electronic-structure calculations. Several strategies to address this challenge include fast multipole \cite{Multipole, Parkkinen2017} and continuous fast multipole methods \cite{ContFastMulti}, hierarchical matrix and block low-rank representations \cite{hier_block_low_rank_tensor, FastMultipoleHierachical}, resolution-of-the-identity (density-fitting) and pivoted Cholesky decomposition techniques \cite{Resolutions, Whitten, Pedersen2024}, and Ewald-type schemes and their particle-mesh variants for periodic systems \cite{Ewald_particle_mesh, SmoothEwald}. These techniques exploit analytic structure, low-rank factorization, locality, or hierarchical decomposition to provide compact representations of the Coulomb operator. Further, efficient finite-difference and multigrid approaches have enabled large-scale electronic-structure calculations on real-space grids.\cite{dogan_solving_2023,briggs_adaptive_2024} 

More recently, stochastic and randomized numerical methods have emerged as complementary approaches. A prominent example is Hutchinson’s method \cite{Hutchinson}, which approximates the trace of a large symmetric matrix using randomized probe vectors. Hutch++ \cite{Hutch++} improves on this approach by first constructing a low-rank deterministic representation $C_{\rm det}$ that captures the dominant spectral components and then applying stochastic estimation to the residual, $C-C_{\rm det}$. This reduces the spectral weight of the residual and consequently the number of stochastic probes required for accurate estimation. Related stochastic resolutions of identity have been developed, including second-order Green’s-function methods for correlation energies and neutral excitations \cite{GreensCorrelation,mejia_stochastic_2023,Hutchinson_for_Greens}. These developments motivate the present work, which constructs a mixed deterministic-stochastic representation of the Coulomb operator based directly on its spectral decomposition.

Related mixed deterministic-stochastic representations of the Coulomb operator have been applied in reciprocal space \cite{StochasticExchange,stochasticRecripocal,planewave}, where the low-$k$ reciprocal lattice vectors are treated deterministically and the high-$k$ components stochastically. In the present work, we instead construct the decomposition in real space based on the spectral range of the Coulomb operator. In particular, we use Chebyshev-filtered subspace iteration (CheFSI) \cite{ChebyshevAccel, CheFSI2006, CheFSI2014} to construct a low-rank deterministic approximation containing the dominant spectral components. The remaining spectral tail is treated stochastically, providing an unbiased correction while preserving the symmetry of the operator.

Importantly, the present approach does not rely on convolution structure, periodicity, or lattice regularity and can be applied to real-space distributions of electronic degrees of freedom, including distorted lattices, molecular geometries, or irregular grids. The resulting operator is benchmarked using a Yukawa-like Coulomb interaction and applied to periodic and perturbed real-space lattices. We further demonstrate its use in self-consistent calculations for the three-dimensional extended Hubbard Hamiltonian. Our results show that this approach provides an accurate and efficient real-space representation of the Coulomb operator, suitable for large-scale electronic-structure simulations.

\section{Methods}
In real space, a Yukawa-like screened Coulomb operator is defined 
\begin{equation}
C(\mathbf{r},\mathbf{r}')
=\frac{e^{-\kappa |\mathbf{r}-\mathbf{r}'|}}
       {|\mathbf{r}-\mathbf{r}'|},
\end{equation}
where $\kappa$ is the inverse screening length, and setting $\kappa=0$ recovers the bare Coulomb interaction. Discretization on a uniform real-space grid produces a dense, symmetric matrix \(C \in \mathbb{R}^{N \times N}\). Our objective is to construct an approximation $C_{\text{approx}}\approx C$,
where operations involving \(C_{\text{approx}}\) scale as \(\mathcal{O}(N r)\) with a controllable rank parameter \(r\), rather than \(\mathcal{O}(N^2)\). Challenges arise from the spectral properties of \(C\): the operator is positive definite, long-ranged, and exhibits a slowly decaying spectrum in non-periodic settings, with dominant components associated with long-range electrostatic interactions. A purely deterministic low-rank truncation fails to capture the extended spectral tail, while a purely stochastic estimator becomes inefficient due to high variance. To address these limitations, we decompose the operator as
\begin{equation}
C = C_{\text{det}} + R,
\end{equation}
where \(C_{\text{det}}\) captures the extremal portion of the spectrum through a deterministic subspace approximation, and \(R\) is the residual containing the broad spectral background. This formulation motivates the mixed deterministic–stochastic representation developed in the following paragraphs, in which Chebyshev-filtered subspaces extract the dominant spectral components while Hutchinson-type probing provides a symmetric, unbiased correction for the remainder of the operator.

\paragraph{Deterministic low-rank component}
To construct the Chebyshev filter, we first estimate the energy window \([a,b]\) using a short Lanczos \cite{GolubVanLoan2013} run with a random start vector,
which defines the scaling coefficients
\[
c = \tfrac{1}{2}(a+b), \qquad d = \tfrac{1}{2}(b-a).
\]
To isolate the largest eigenmodes, we define a scaled operator whose eigenvalues lie within the interval $[-1, 1]$, 
\[
\tilde{C}v = - \frac{C v - c v}{d},
\]
where we take the negative of the operator to target the largest eigenmodes. Starting from a random matrix sampled from a Gaussian distribution \( Q_0 \in \mathbb{R}^{N\times r} \), we orthonormalize this matrix. Then we apply the block Chebyshev recurrence
\[
Y_{k+1} = 2\,\tilde{C} Y_k - Y_{k-1}, \qquad k=1,\ldots,\text{deg},
\]
with \(Y_0 = Q_0\) and \(Y_1 = \tilde{C} Q_0\) \cite{OGCheby,CheFSI2006, Chen2025}.
After each outer iteration, the resulting block is reorthonormalized using Gram-Schmidt orthogonalization, i.e., $
Q \leftarrow \operatorname{orth}(Y_{\text{deg}})$. Repeated filtering concentrates \(Q\) within the desired spectral subspace, providing an approximate invariant basis of dimension \(r\). The corresponding projected matrix is
\[
B = Q^\top C Q, \qquad
C_{\text{det}} = Q B Q^\top,
\]
which forms the deterministic low-rank approximation that captures the well-resolved, extreme portion of the spectrum.

\paragraph{Stochastic tail correction}
To recover contributions from the remainder of the spectrum,
we estimate the residual operator \( R = C - C_{\text{det}} \) via randomized probing. Let \( r_m \in \{\pm 1\}^N \) be random probe vectors, and define
\[
e_m = R r_m = (C - Q B Q^\top) r_m.
\]
An unbiased, symmetric estimator of the residual contribution is then
\[
\Delta C = \frac{1}{2M} \sum_{m=1}^M
\big( r_m e_m^\top + e_m r_m^\top \big).
\]
Combining the deterministic low-rank component with the stochastic residual correction produces, $C_{\text{approx}}
= Q B Q^\top
+ \Delta C$, which remains symmetric by construction and unbiased in expectation. The first term resolves the low-rank subspace deterministically, while the second term captures the stochastic correction representing the unresolved tail of the spectrum.

\paragraph{Jackknife bias correction}
Although the stochastic approximation of the Coulomb operator is unbiased in expectation, its use within the nonlinear self-consistent field (SCF) procedure can introduce a finite-sample bias in observables such as total energies and orbital eigenvalues. To reduce this bias, we apply a jackknife correction to the ensemble-averaged quantities, following the prescription of Ref. \cite{savchenko_bias_2026}.

Let $\{E_i\}_{i=1}^{N_{\mathrm{stoc}}}$ denote the total energies obtained from $N_{\mathrm{stoc}}$ independent stochastic realizations of the Coulomb operator. Here, $N_{\rm stoc}$ denotes the number of independent stochastic realizations and is distinct from the number of probe vectors $M$ used in constructing each Coulomb approximation. The full-sample estimator is
\[
E(N_{\mathrm{stoc}}) 
= \frac{1}{N_{\mathrm{stoc}}} \sum_{i=1}^{N_{\mathrm{stoc}}} E_i .
\]
Following Ref.~\cite{savchenko_bias_2026}, we construct a bias-corrected estimator using stochastic ensembles of different sizes. Let
\[
E\!\left(\frac{N_{\mathrm{stoc}}}{2}\right),
\qquad
E\!\left(\frac{N_{\mathrm{stoc}}}{2}+1\right),
\]
denote ensemble-averaged energies computed using independent stochastic samples. The jackknife-corrected (JK) estimator is then defined as
\[
E_{\mathrm{JK}}(N_{\mathrm{stoc}})
=
2E(N_{\mathrm{stoc}})
-
\frac{1}{2}
\left[
E\!\left(\frac{N_{\mathrm{stoc}}}{2}\right)
+
E\!\left(\frac{N_{\mathrm{stoc}}}{2}+1\right)
\right].
\]

This construction reduces the leading-order bias associated with finite stochastic sample size while preserving the expected $O(N_{\mathrm{stoc}}^{-1/2})$ convergence of the residual statistical error. In all reported results, jackknife-corrected estimators are used for stochastic observables obtained from SCF calculations.

\section{\label{sec:Results}Results}

We benchmark the approximate Coulomb operator against an exact, fully discretized treatment within self-consistent field (SCF) calculations using an extended Hubbard Hamiltonian:
\begin{equation*}
    H \;=\; - \sum_{i\neq j,\sigma} t_{ij}\, c^{\dagger}_{i\sigma} c_{j\sigma}
    \;-\; \mu \sum_i n_i 
    \;+\; U \sum_i n_{i\uparrow} n_{i\downarrow}
    \;+\; \frac{1}{2}\sum_{i\neq j} V_{ij}\, n_i n_j \,.
\end{equation*}
Here, $c_{i\sigma}^{\dagger}$ and $c_{i\sigma}$ are the respective creation and annihilation operators for an electron with spin $\sigma$ on lattice site $i$; $t_{ij}$ is the lattice hopping between sites $i$ and $j$; $n_{i\sigma}=c_{i\sigma}^{\dagger}c_{i\sigma}$ and $n_i=\sum_{\sigma}n_{i\sigma}$ are the spin-resolved and total site occupations; $\mu$ is the chemical potential; and $U$ is the on-site interaction strength. The Coulomb interaction is
\begin{equation*}
    V_{ij} = \begin{cases}
      \dfrac{e^{-\kappa |\mathbf r_i-\mathbf r_j|}}{|\mathbf r_i-\mathbf r_j|}\,, & i\neq j,\\[8pt]
      0\,, & i=j\,.
    \end{cases}
\end{equation*}
For all calculations, the hopping parameter was set to $t_{ij}=1$ and the inverse screening length is set to $\kappa=2$.

Following SCF convergence, the occupied orbital energies produced using the approximate operator are compared against those from the exact operator, with accuracy quantified by the maximum relative error:
\[
\varepsilon
=
\max_{1 \le i \le N_{\mathrm{occ}}}
\frac{
| \lambda_i^{\text{approx}} - \lambda_i^{\text{exact}} |
}{
 | \lambda_i^{\text{exact}} | + \delta
},
\]
where $\lambda_i^{\rm approx}$ and $\lambda_i^{\rm exact}$ are the occupied orbital energies obtained using the approximate and exact Coulomb operators, respectively, and $N_{\rm occ}$ is the number of occupied states.

\begin{figure}[htbp]
    \centering
    \includegraphics[width=0.7\linewidth]{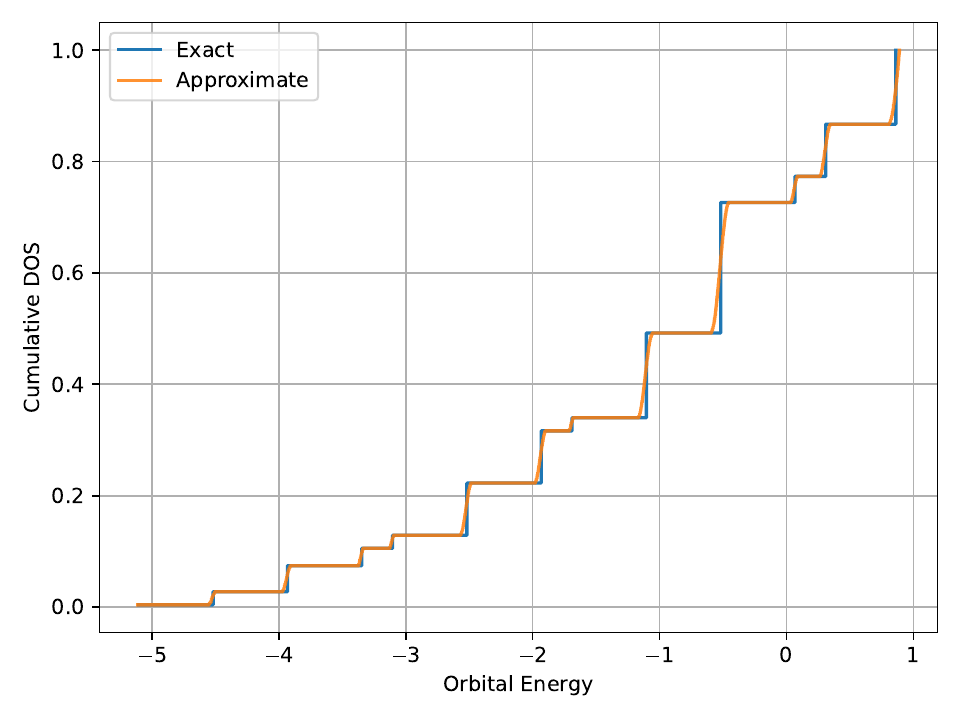}
    \caption{Cumulative Density of States (cDOS) comparison for a uniform lattice with $N=1000$ grid points, computed using $M=200$ stochastic probes and $r=70$ deterministic eigenvectors.}
    \label{fig:cdos_L6}
\end{figure}

We first compare the cumulative density of states (cDOS) obtained using the approximate and exact Coulomb operators. Fig. \ref{fig:cdos_L6} shows that the approximate operator accurately reproduces the global spectral structure of the exact operator. The cDOS curves track each other across the entire spectrum, indicating that the low-rank plus stochastic representation preserves the dominant features of the Coulomb interaction. 

\begin{figure}[htbp]
    \centering
    \begin{subfigure}{0.48\linewidth}
        \centering
        \includegraphics[width=\linewidth]{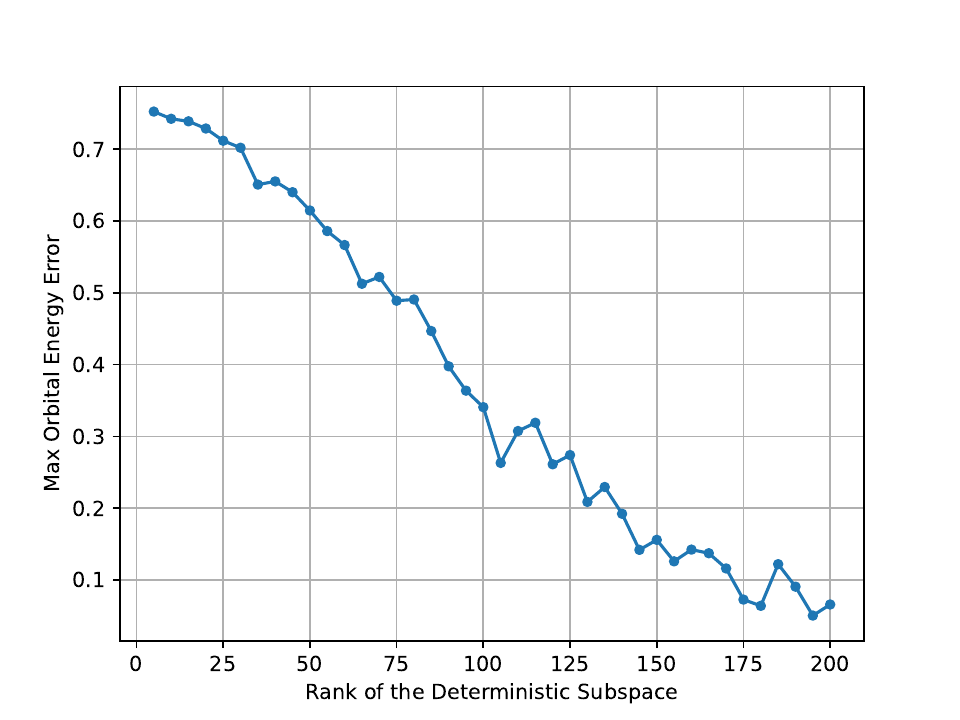}
        \caption{Error vs deterministic eigenvectors $r$ (fixed stochastic probes $M=100$).}
        \label{fig:error_rank}
    \end{subfigure}
    \hfill
    \begin{subfigure}{0.48\linewidth}
        \centering
        \includegraphics[width=\linewidth]{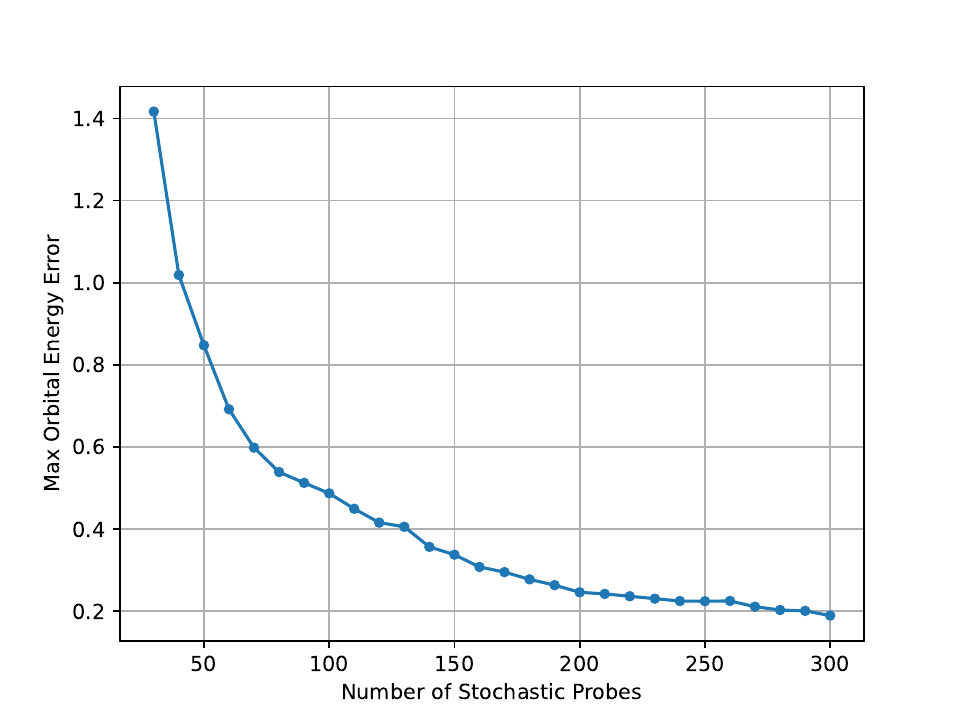}
        \caption{Error vs stochastic probes $M$ (fixed deterministic subspace rank $r=40$).}
        \label{fig:error_probes}
    \end{subfigure}
    
    \caption{Convergence of the Coulomb operator approximation. (a) Relative error of occupied orbital energies as a function of deterministic subspace rank $r$ at fixed $M=100$ and system size $N=1000$. (b) Relative error as a function of the number of stochastic probes $M$ at fixed $r=40$.}
    \label{fig:error_combined}
\end{figure}

We next examine convergence with respect to the deterministic subspace rank \(r\) and the number of stochastic probes \(M\). In Fig. \ref{fig:error_rank}, the error decreases monotonically with $r$, following an approximate \(1/r\) behavior. At fixed rank $r=40$, Fig. \ref{fig:error_probes} shows that the error decays proportionally as \(M^{-1/2}\), reflecting the Monte Carlo nature of the Hutch++ estimator.

\begin{figure}[htbp]
    \centering
    \begin{subfigure}{0.48\linewidth}
        \centering
        \includegraphics[width=\linewidth]{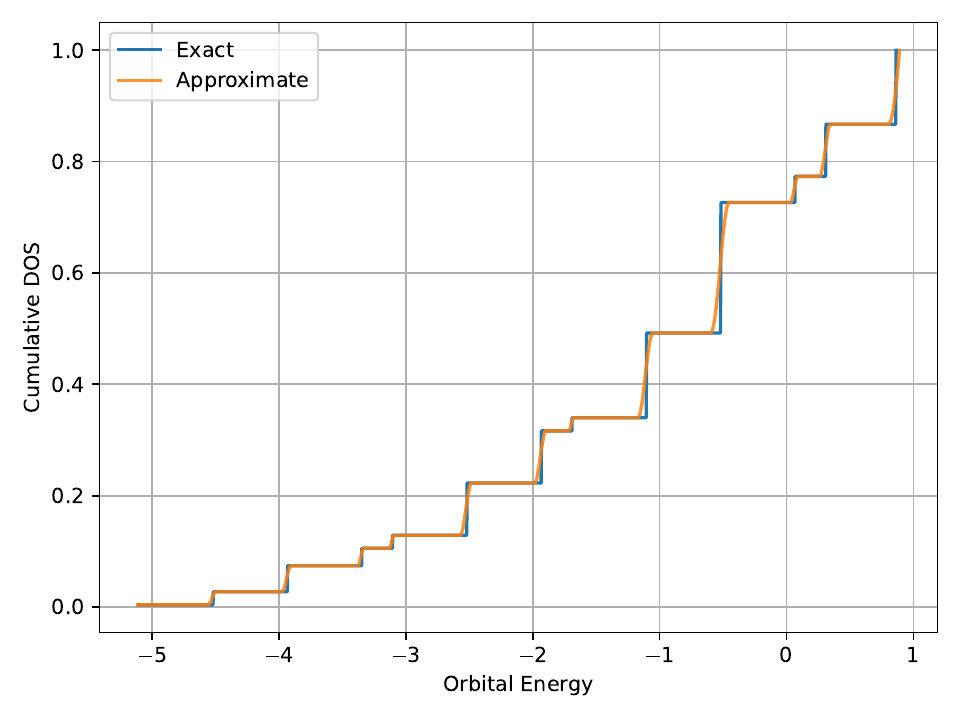}
        \caption{Perturbed periodic lattice.}
        \label{fig:cdos_perturbed}
    \end{subfigure}
    \hfill
    \begin{subfigure}{0.48\linewidth}
        \centering
        \includegraphics[width=\linewidth]{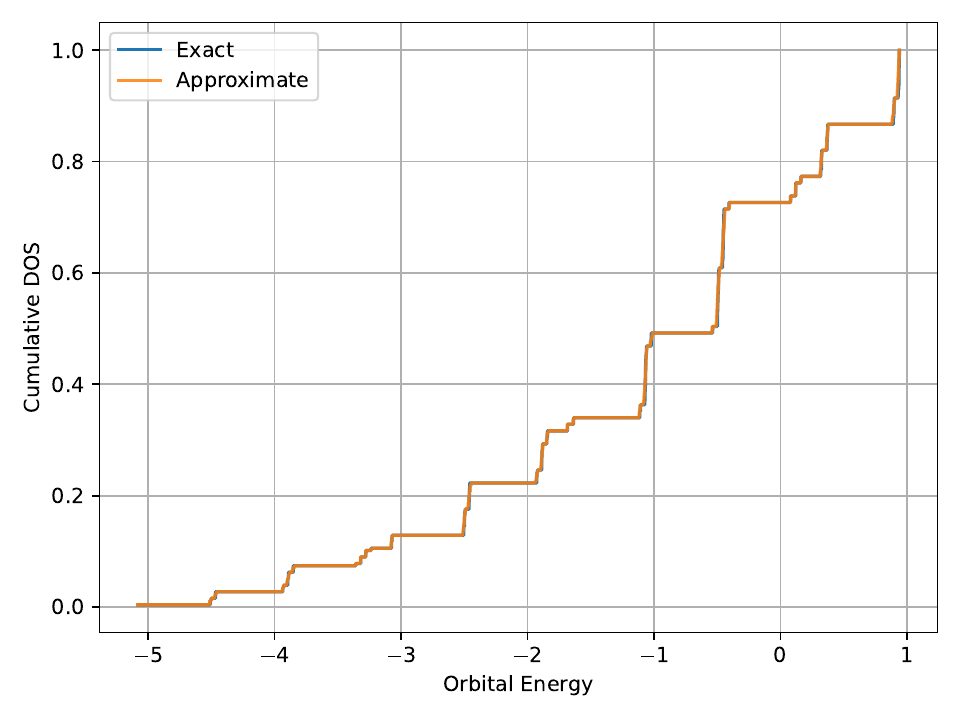}
        \caption{Non-periodic perturbed lattice.}
        \label{fig:cdos_nonperiodic}
    \end{subfigure}

    \caption{cDOS obtained for (a) Perturbed periodic lattice and (b) perturbed lattice with open boundary conditions. Results are shown for $M=1000$ stochastic probes with deterministic eigenvectors $r=30$ and $r=70$, respectively.}
    \label{fig:cdos_combined}
\end{figure}

To examine the effect of lattice distortions, each site was displaced by a fixed fraction of the lattice spacing. The resulting cDOS is shown in Fig.~\ref{fig:cdos_perturbed}. The corresponding open-boundary calculation is shown in Fig.~\ref{fig:cdos_nonperiodic}. In both cases, the approximate and exact cDOS remain in close agreement.

\begin{figure}[htbp]
    \centering
    \includegraphics[width=0.7\linewidth]{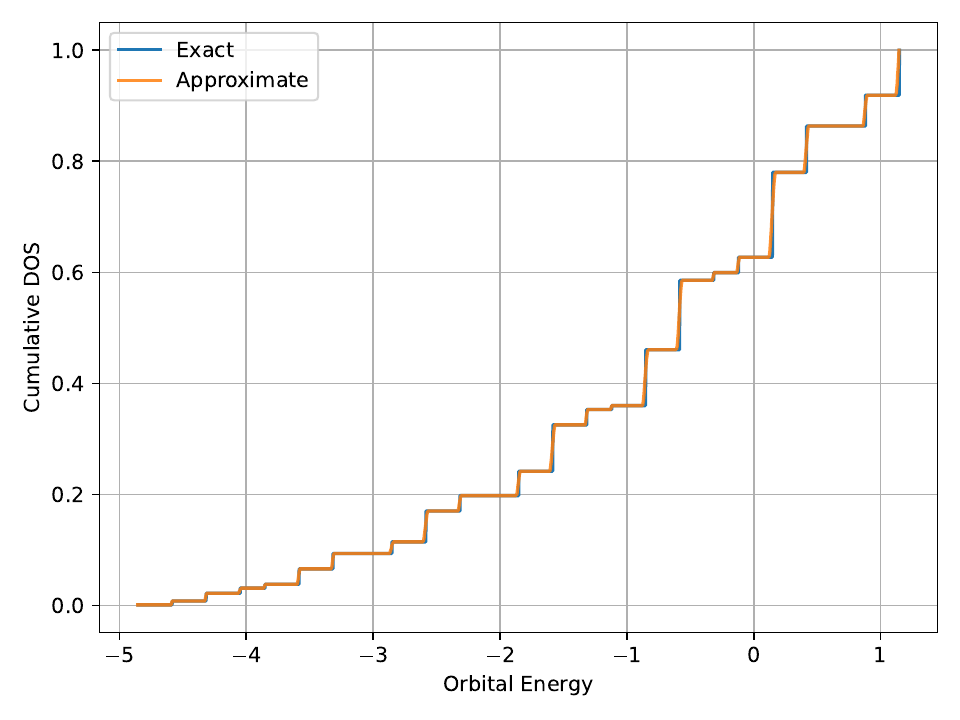}
    \caption{cDOS for a larger unperturbed periodic lattice with lattice size $L=10$.}
    \label{fig:cdos_L10}
\end{figure}

Fig. \ref{fig:cdos_L10} shows the cDOS for the larger unperturbed periodic lattice. The approximate and exact spectra remain in close agreement. For a fixed level of accuracy, the number of stochastic probe vectors grows more slowly than the total number of grid points.

\begin{figure}[htbp]
    \centering
    \includegraphics[width=0.7\linewidth]{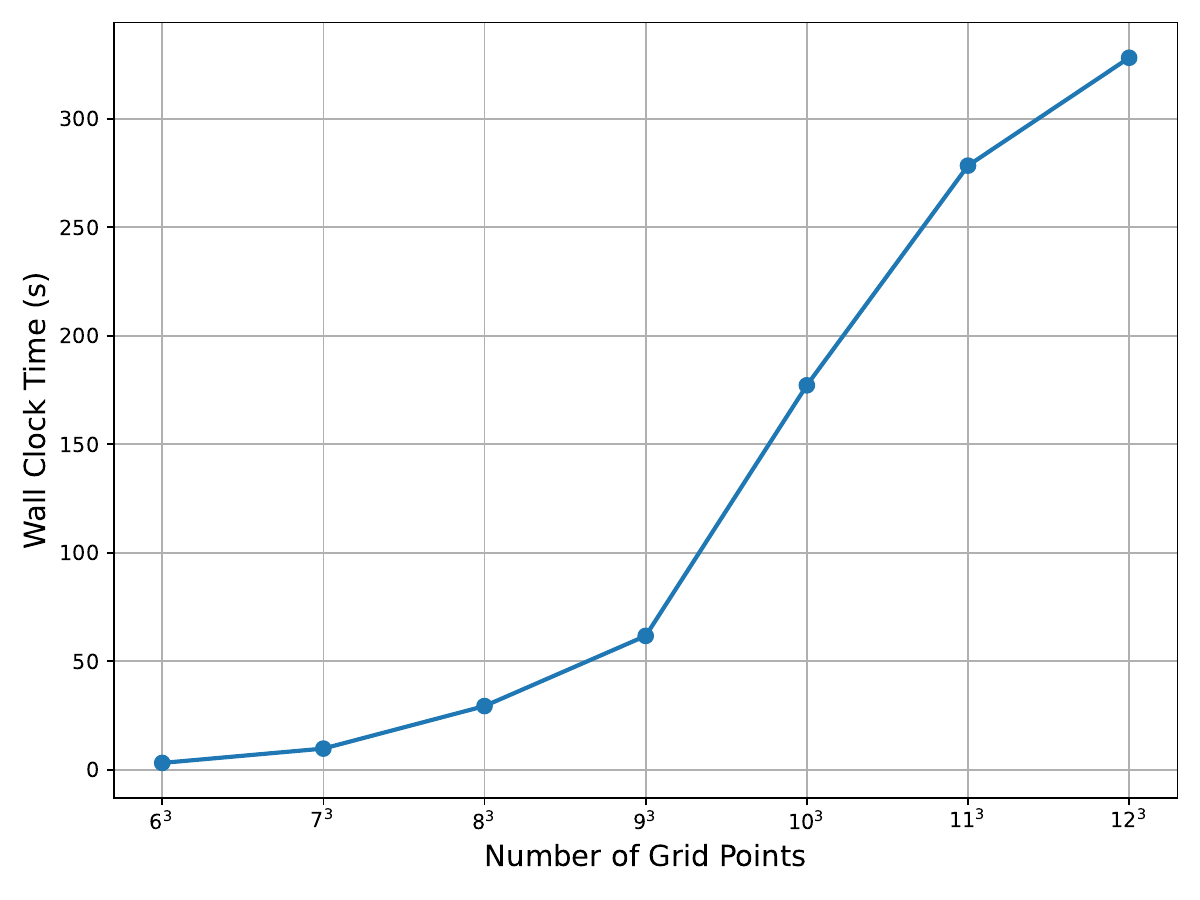}
    \caption{Wall-clock time required for initialization to achieve a relative error $\le 0.001$ as a function of the number of grid points $L^3$.}
    \label{fig:scaling}
\end{figure}

The computational cost required to reach a relative error of \(0.001\) is shown in Fig.~\ref{fig:scaling}. Although the runtime grows superlinearly with system length \(L\), the scaling is close to linear in the total number of grid points $N$, consistent with expectations for Hutch++ kernel sketching in a serial implementation.

\section{Conclusions\label{sec:conclusion}}

We have presented a mixed deterministic-stochastic strategy for constructing an efficient real-space representation of the Coulomb operator that is compatible with electronic-structure calculations across a wide range of chemically relevant settings. The method combines a Chebyshev-filtered subspace procedure, which captures the dominant long-range interaction structure in a compact low-rank form, with an unbiased stochastic correction that reconstructs the remaining non-dominant components of the interaction. This decomposition provides controllable accuracy: increasing the deterministic rank systematically improves the representation of the long-range modes, while the stochastic estimator converges with the expected \(\mathcal{O}(M^{-1/2})\) behavior as the number of probes increases.

Numerical tests on periodic, perturbed, and non-periodic real-space grids demonstrate that the method accurately reproduces the occupied orbital energies and cumulative density of states obtained from self-consistent-field calculations using the fully deterministic Coulomb operator. The observed error reductions follow the predicted trends with respect to deterministic rank and stochastic probe count, and the method remains robust under geometric distortions that break lattice regularity. The overall computational effort scales as \(\mathcal{O}(rN + MN)\), with an initialization cost of $\mathcal{O}(rN^2 + MN^2)$.

The present formulation provides a flexible real-space representation of the Coulomb operator without requiring convolution structure, periodicity, or lattice regularity. Future work will apply the method to ground-state DFT calculations in real space, beginning with evaluation of the Hartree term. Extension of the deterministic-stochastic representation to the nonlocal exact-exchange operator will then be explored for large-scale generalized Kohn-Sham DFT calculations. Further, parallelized implementations of the CheFSI and stochastic-probing stages will enable these applications on substantially larger real-space grids.

\section*{Acknowledgments}
We thank Mykola Sereda for useful discussions. This work is supported by the National Science Foundation (NSF) under Grant No. CHE-2245253 and U.S-Israel Binational Science Foundation (BSF) under Grant No. 2024265. 

\section*{Declaration of Competing Interest}
The authors declare that they have no known competing financial interests or personal relationships that could have appeared to influence the work reported in this paper.

\section*{Data Availability Statement}
The data that support the findings of this study are available from the corresponding author upon reasonable request.

\pagebreak

\bibliographystyle{unsrt}
\bibliography{bib}

@article{mejia_stochastic_2023,
	title = {Stochastic {Real}-{Time} {Second}-{Order} {Green}’s {Function} {Theory} for {Neutral} {Excitations} in {Molecules} and {Nanostructures}},
	volume = {19},
	copyright = {https://doi.org/10.15223/policy-029},
	issn = {1549-9618, 1549-9626},
	doi = {10.1021/acs.jctc.3c00296},
	number = {16},
	urldate = {2026-08-31},
	journal = {Journal of Chemical Theory and Computation},
	author = {Mejía, Leopoldo and Yin, Jia and Reichman, David R. and Baer, Roi and Yang, Chao and Rabani, Eran},
	month = aug,
	year = {2023},
	pages = {5563--5571},
}

@article{dogan_solving_2023,
	title = {Solving the electronic structure problem for over 100 000 atoms in real space},
	volume = {7},
	issn = {2475-9953},
	doi = {10.1103/PhysRevMaterials.7.L063001},
	number = {6},
	urldate = {2026-08-31},
	journal = {Physical Review Materials},
	author = {Dogan, Mehmet and Liou, Kai-Hsin and Chelikowsky, James R.},
	month = jun,
	year = {2023},
	pages = {L063001},
}

@article{briggs_adaptive_2024,
	title = {Adaptive finite differencing in high accuracy electronic structure calculations},
	volume = {10},
	issn = {2057-3960},
	doi = {10.1038/s41524-024-01203-y},
	number = {1},
	urldate = {2026-08-31},
	journal = {npj Computational Materials},
	author = {Briggs, E. L. and Lu, Wenchang and Bernholc, J.},
	month = jan,
	year = {2024},
	pages = {17},
}

@book{GolubVanLoan2013,
  author    = {Gene H. Golub and Charles F. Van Loan},
  title     = {Matrix Computations},
  edition   = {4th},
  publisher = {Johns Hopkins University Press},
  year      = {2013},
  address   = {Baltimore, MD}
}

@article{LinearExact,
    author = {Ochsenfeld, Christian and White, Christopher A. and Head-Gordon, Martin},
    title = {Linear and Sublinear Scaling Formation of {Hartree--Fock}-Type Exchange Matrices},
    journal = {The Journal of Chemical Physics},
    volume = {109},
    number = {5},
    pages = {1663--1669},
    year = {1998},
    month = aug,
    doi = {10.1063/1.476741},
    url = {https://doi.org/10.1063/1.476741}
}

@book{martin2004electronic,
  title     = {Electronic Structure: Basic Theory and Practical Methods},
  author    = {Martin, Richard M.},
  year      = {2004},
  publisher = {Cambridge University Press},
  address   = {Cambridge}
}

@article{CheFSI2006,
  title = {Self-Consistent-Field Calculations Using {Chebyshev}-Filtered Subspace Iteration},
  journal = {Journal of Computational Physics},
  volume = {219},
  number = {1},
  pages = {172--184},
  year = {2006},
  issn = {0021-9991},
  doi = {10.1016/j.jcp.2006.03.017},
  url = {https://www.sciencedirect.com/science/article/pii/S002199910600146X},
  author = {Yunkai Zhou and Yousef Saad and Murilo L. Tiago and James R. Chelikowsky}
}

@article{CheFSI2014,
  title = {{Chebyshev}-Filtered Subspace Iteration Method Free of Sparse Diagonalization for Solving the {Kohn--Sham} Equation},
  journal = {Journal of Computational Physics},
  volume = {274},
  pages = {770--782},
  year = {2014},
  issn = {0021-9991},
  doi = {10.1016/j.jcp.2014.06.056},
  url = {https://www.sciencedirect.com/science/article/pii/S0021999114004744},
  author = {Yunkai Zhou and James R. Chelikowsky and Yousef Saad}
}

@article{stochasticRecripocal,
  author = {Sereda, Mykola and Allen, Tucker and Bradbury, Nadine C. and Ibrahim, Khaled Z. and Neuhauser, Daniel},
  title = {Sparse-Stochastic Fragmented Exchange for Large-Scale Hybrid Time-Dependent Density Functional Theory Calculations},
  journal = {Journal of Chemical Theory and Computation},
  volume = {20},
  number = {10},
  pages = {4196--4204},
  year = {2024},
  doi = {10.1021/acs.jctc.4c00260},
  note = {PMID: 38713513},
  url = {https://doi.org/10.1021/acs.jctc.4c00260}
}

@article{ChebyshevAccel,
  author = {Youcef Saad},
  title = {Chebyshev Acceleration Techniques for Solving Nonsymmetric Eigenvalue Problems},
  journal = {Mathematics of Computation},
  volume = {42},
  number = {166},
  pages = {567--588},
  year = {1984},
  publisher = {American Mathematical Society},
  url = {http://www.jstor.org/stable/2007602}
}

@article{StochasticExchange,
  author = {Bradbury, Nadine C. and Allen, Tucker and Nguyen, Minh and Neuhauser, Daniel},
  title = {Deterministic/Fragmented-Stochastic Exchange for Large-Scale Hybrid {DFT} Calculations},
  journal = {Journal of Chemical Theory and Computation},
  volume = {19},
  number = {24},
  pages = {9239--9247},
  year = {2023},
  doi = {10.1021/acs.jctc.3c00987},
  note = {PMID: 38051791},
  url = {https://doi.org/10.1021/acs.jctc.3c00987}
}

@article{ContFastMulti,
  title = {The Continuous Fast Multipole Method},
  journal = {Chemical Physics Letters},
  volume = {230},
  number = {1},
  pages = {8--16},
  year = {1994},
  issn = {0009-2614},
  doi = {10.1016/0009-2614(94)01128-1},
  url = {https://www.sciencedirect.com/science/article/pii/0009261494011281},
  author = {Christopher A. White and Benny G. Johnson and Peter M.W. Gill and Martin Head-Gordon}
}

@article{Hutchinson_for_Greens,
  author = {Mejía, Leopoldo and Sharma, Sandeep and Baer, Roi and Chan, Garnet Kin-Lic and Rabani, Eran},
  title = {Convergence Analysis of the Stochastic Resolution of Identity: Comparing {Hutchinson} to {Hutch++} for the Second-Order {Green's} Function},
  journal = {Journal of Chemical Theory and Computation},
  volume = {20},
  number = {17},
  pages = {7494--7502},
  year = {2024},
  doi = {10.1021/acs.jctc.4c00862},
  note = {PMID: 39189663},
  url = {https://doi.org/10.1021/acs.jctc.4c00862}
}

@inproceedings{Hutch++,
  author = {Raphael A. Meyer and Cameron Musco and Christopher Musco and David P. Woodruff},
  title = {{Hutch++}: Optimal Stochastic Trace Estimation},
  booktitle = {2021 Symposium on Simplicity in Algorithms (SOSA)},
  pages = {142--155},
  year = {2021},
  doi = {10.1137/1.9781611976496.16},
  url = {https://epubs.siam.org/doi/abs/10.1137/1.9781611976496.16}
}

@article{hier_block_low_rank_tensor,
  author = {Xing, Xin and Huang, Hua and Chow, Edmond},
  title = {A Linear Scaling Hierarchical Block Low-Rank Representation of the Electron Repulsion Integral Tensor},
  journal = {The Journal of Chemical Physics},
  volume = {153},
  number = {8},
  pages = {084119},
  year = {2020},
  month = aug,
  doi = {10.1063/5.0010732},
  url = {https://doi.org/10.1063/5.0010732}
}

@article{Hutchinson,
  author = {M.F. Hutchinson},
  title = {A Stochastic Estimator of the Trace of the {Laplacian} Smoothing Splines},
  journal = {Communications in Statistics - Simulation and Computation},
  volume = {19},
  number = {2},
  pages = {433--450},
  year = {1990},
  publisher = {Taylor \& Francis},
  doi = {10.1080/03610919008812866},
  url = {https://doi.org/10.1080/03610919008812866}
}

@article{Multipole,
  author = {Toivanen, Elias A. and Losilla, Sergio A. and Sundholm, Dage},
  title = {The Grid-Based Fast Multipole Method -- A Massively Parallel Numerical Scheme for Calculating Two-Electron Interaction Energies},
  journal = {Phys. Chem. Chem. Phys.},
  year = {2015},
  volume = {17},
  number = {47},
  pages = {31480--31490},
  publisher = {The Royal Society of Chemistry},
  doi = {10.1039/C5CP01173F},
  url = {http://dx.doi.org/10.1039/C5CP01173F}
}

@article{Ewald_particle_mesh,
  author = {Kawata, Masaaki and Mikami, Masuhiro and Nagashima, Umpei},
  title = {Computationally Efficient Method to Calculate the {Coulomb} Interactions in Three-Dimensional Systems with Two-Dimensional Periodicity},
  journal = {The Journal of Chemical Physics},
  volume = {116},
  number = {8},
  pages = {3430--3448},
  year = {2002},
  month = feb,
  doi = {10.1063/1.1445103},
  url = {https://doi.org/10.1063/1.1445103}
}

@article{Resolutions,
  title = {Resolutions of the {Coulomb} Operator: {II}. The {Laguerre} Generator},
  journal = {Chemical Physics},
  volume = {356},
  number = {1},
  pages = {86--90},
  year = {2009},
  note = {Moving Frontiers in Quantum Chemistry},
  issn = {0301-0104},
  doi = {10.1016/j.chemphys.2008.10.047},
  url = {https://www.sciencedirect.com/science/article/pii/S0301010408004862},
  author = {Peter M.W. Gill and Andrew T.B. Gilbert}
}

@article{GreensCorrelation,
  author = {Neuhauser, Daniel and Baer, Roi and Zgid, Dominika},
  title = {Stochastic Self-Consistent Second-Order {Green's} Function Method for Correlation Energies of Large Electronic Systems},
  journal = {Journal of Chemical Theory and Computation},
  volume = {13},
  number = {11},
  pages = {5396--5403},
  year = {2017},
  doi = {10.1021/acs.jctc.7b00792},
  note = {PMID: 28961398},
  url = {https://doi.org/10.1021/acs.jctc.7b00792}
}

@article{planewave,
  title = {Efficient Plane-Wave Approach to Generalized {Kohn--Sham} Density Functional Theory of Solids with Mixed Deterministic and Stochastic Exchange},
  author = {Allen, Tucker and Li, Barry Y. and Duong, Tim and Williams, Kajsa and Neuhauser, Daniel},
  journal = {Phys. Rev. B},
  volume = {112},
  number = {15},
  pages = {155104},
  year = {2025},
  month = oct,
  publisher = {American Physical Society},
  doi = {10.1103/PhysRevB.112.155104},
  url = {https://link.aps.org/doi/10.1103/PhysRevB.112.155104}
}

@article{savchenko_bias_2026,
  title = {Bias and Its Control in Stochastic Approaches to Electronic-Structure Theory},
  author = {Savchenko, Pavel and Adhikari, Sayak and Hadad, Efrat and Rabani, Eran and Baer, Roi},
  journal = {Journal of Chemical Theory and Computation},
  year = {2026},
  month = mar,
  doi = {10.1021/acs.jctc.5c01970},
  url = {https://pubs.acs.org/doi/10.1021/acs.jctc.5c01970}
}

@inproceedings{FastMultipoleHierachical,
  author = {Yokota, Rio and Ibeid, Huda and Keyes, David},
  editor = {Sakurai, Tetsuya and Zhang, Shao-Liang and Imamura, Toshiyuki and Yamamoto, Yusaku and Kuramashi, Yoshinobu and Hoshi, Takeo},
  title = {Fast Multipole Method as a Matrix-Free Hierarchical Low-Rank Approximation},
  booktitle = {Eigenvalue Problems: Algorithms, Software and Applications in Petascale Computing},
  publisher = {Springer International Publishing},
  address = {Cham},
  year = {2017},
  pages = {267--286},
  isbn = {978-3-319-62426-6}
}

@article{Parkkinen2017,
  author = {Parkkinen, Pauli and Losilla, Sergio A. and Solala, Eelis and Toivanen, Elias A. and Xu, Wen-Hua and Sundholm, Dage},
  title = {A Generalized Grid-Based Fast Multipole Method for Integrating {Helmholtz} Kernels},
  journal = {Journal of Chemical Theory and Computation},
  volume = {13},
  number = {2},
  pages = {654--665},
  year = {2017},
  doi = {10.1021/acs.jctc.6b01207}
}

@article{Whitten,
  author = {Whitten, J. L.},
  title = {{Coulombic} Potential Energy Integrals and Approximations},
  journal = {The Journal of Chemical Physics},
  volume = {58},
  number = {10},
  pages = {4496--4501},
  year = {1973},
  month = may,
  doi = {10.1063/1.1679012},
  url = {https://doi.org/10.1063/1.1679012}
}

@article{SmoothEwald,
  author = {Essmann, Ulrich and Perera, Lalith and Berkowitz, Max L. and Darden, Tom and Lee, Hsing and Pedersen, Lee G.},
  title = {A Smooth Particle Mesh {Ewald} Method},
  journal = {The Journal of Chemical Physics},
  volume = {103},
  number = {19},
  pages = {8577--8593},
  year = {1995},
  month = nov,
  doi = {10.1063/1.470117},
  url = {https://doi.org/10.1063/1.470117}
}

@article{OGCheby,
  author = {Tal-Ezer, H. and Kosloff, R.},
  title = {An Accurate and Efficient Scheme for Propagating the Time-Dependent {Schr\"odinger} Equation},
  journal = {The Journal of Chemical Physics},
  volume = {81},
  number = {9},
  pages = {3967--3971},
  year = {1984},
  month = nov,
  doi = {10.1063/1.448136},
  url = {https://doi.org/10.1063/1.448136}
}

@article{Chen2025,
  author = {Chen, Kyle and Li, Barry Y. and Allen, Tucker and Neuhauser, Daniel},
  title = {Mixed Plane-Wave and Localized Orbital Basis for Sparse-Stochastic Hybrid {Time-Dependent Density Functional Theory}},
  journal = {Journal of Chemical Theory and Computation},
  year = {2025},
  volume = {21},
  number = {17},
  pages = {8509--8517},
  doi = {10.1021/acs.jctc.5c01025},
  pmid = {40833030}
}

@article{Pedersen2024,
  title = {The versatility of the {Cholesky} decomposition in electronic structure theory},
  author = {Pedersen, Thomas Bondo and Lehtola, Susi and Fdez. Galv{\'a}n, Ignacio and Lindh, Roland},
  journal = {WIREs Computational Molecular Science},
  volume = {14},
  number = {1},
  pages = {e1692},
  year = {2024},
  doi = {10.1002/wcms.1692}
}

\end{document}